\documentclass[prl,aps,twocolumn,eqsecnum,showpacs]{revtex4}

\usepackage{dcolumn}
\usepackage{amsmath}
\usepackage{graphics}

\begin{document}
\title{Imaging-based Parametric Resonance in an Optical Dipole Atom Trap}

\author{S. Balik, A.L. Win and M.D. Havey}
\affiliation{Department of Physics, Old Dominion University,
Norfolk, VA 23529}
\date{\today }
\email{mhavey@odu.edu}

\date{\today }

\begin{abstract}
We report sensitive detection of parametric resonances in a
high-density sample of ultracold $^{87}Rb$ atoms confined to a
far-off-resonance optical dipole trap.  Fluorescence imaging of
the expanded ultracold atom cloud after a period of parametric
excitation shows significant modification of the atomic spatial
distribution and has high sensitivity compared with traditional
measurements of parametrically-driven trap loss. Using this
approach, a significant shift of the parametric resonance
frequency is observed, and attributed to the anharmonic shape of
the dipole trap potential.
\end{abstract}

\maketitle%

Parametrically driven processes are ubiquitous in nature, and
occur in a range of classical and quantum systems
\cite{Pippard,Strogatz,Davidson,Insperger,Kobes}. In a
parametrically driven system, a system parameter may be
harmonically varied in time. For a one-dimensional harmonic
oscillator with a characteristic frequency $\omega_{o}$, such
excitation generates resonances at frequencies $\omega_{p} =
\frac{2\omega_{o}}{n}$, where n is an integer n = 1,2,.. A
familiar example is the inverted pendulum, which may be stabilized
against decay of small oscillations by appropriately driving the
pivot point. One important role of parametric excitation in atomic
physics is its influence on ultracold atoms confined in
magneto-optical \cite{Kim1}, magnetic
\cite{Kumakura,Yanbo,Zhang,Singh}, and optical dipole traps
\cite{Friebel,Savard,Gehm,Grimm,Gardiner}. For situations where
long lived traps are desirable, excitation due to noise in the
trap parameters normally heats the atoms and thus leads to atoms
being expelled from the trap. Under some circumstances parametric
driving of a system with an anharmonic potential can also cool the
atom cloud \cite{Kumakura,Yanbo,Zhang}. The lifetime of the trap
is limited by technical noise in any of the trap parameters
\cite{Savard,Gehm,Grimm}, two important ones for optical dipole
traps being the pointing stability and the intensity noise
characteristics of the trapping lasers.  In addition to this
deleterious effect, parametric resonance is a useful and widely
employed tool for characterizing the shape of the trap, including
the harmonic frequencies of the trap \cite{Grimm}. This is
important in experiments where the spatial shape of the trapping
potential is needed for interpretation of the results. In
experiments in our laboratory, we are interested in obtaining a
high density of atoms in an optical dipole trap, and in knowing
the peak density as well as possible. This requires that the
number of atoms, atom temperature, and shape of the trapping
potential be well known.  A main goal of these experiments is to
study the electromagnetic analog of Anderson Localization
\cite{Havey1,CommentAMO,LPLReview,Akkermans1} of light in an
ultracold atomic gas.

In this paper we present experimental results on parametric
excitation of a sample of ultracold $^{87}Rb$ atoms confined in an
optical dipole trap. Such studies are frequently done by measuring
the relative number of atoms that are removed from the trap as
excitation parameters are varied.  In the present case we measure
instead modification of images of the ultracold cloud following
parametric excitation and a subsequent period of free expansion.
Analysis of the images provides a more sensitive measure of the
effects of parametric excitation, including clear evidence of the
frequency shift of the parametric resonance frequency due to trap
anharmonicity \cite{Zhang,Yanbo,Jauregui}. In the following we
provide a brief description of the experimental apparatus. This is
followed by presentation of the measured images and their
analysis.

\begin{figure}[htpb]
\includegraphics{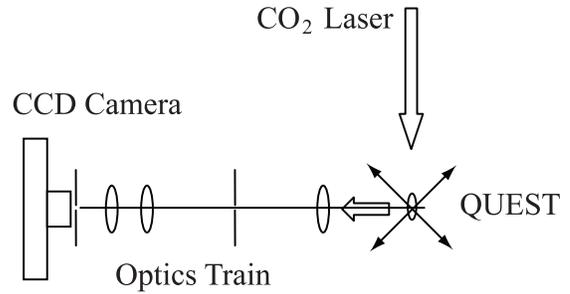}%
\caption{Schematic drawing of the experimental apparatus. In the
figure QUEST stands for quasi electrostatic trap, and CCD
represents charge coupled device. Drawing not to scale.}
\label{fig1}%
\end{figure}

A schematic of the experimental apparatus is shown in Fig. 1. In
the figure, the central part of the experimental apparatus is a
magneto optical trap (MOT) confining ultra cold $^{87}Rb$ atoms.
The MOT is a vapor-loaded trap formed in a vacuum chamber with a
base pressure $\sim 10^{-9}$ Torr. The six MOT beams are derived
from an external cavity diode laser (ECDL) with the grating
arranged in Littrow configuration. The master diode laser is
frequency locked to a saturation absorption feature produced in a
Rb vapor cell. The laser power is increased by injecting the
output into a slave laser, thus providing more than 20 mW of light
in trapping laser beams of cross sectional area $\sim$ 2 $cm^{2}$.
The slave output is switched and spectrally shifted as required
with an acousto optical modulator (AOM) to a frequency about 18
MHz below the $^{87}Rb$ $F=2\to F'=3$ trapping transition.  The
repumper laser is also an ECDL of the same design as the MOT
laser, and is locked to the $F=1\to F'=2$ hyperfine transition.
The repumper delivers a beam of maximum intensity $\sim$ 0.6
$mW/cm^{2}$ and is delivered along the same optical path as the
trapping laser beams. Repumper switching is controlled with an
AOM.

The cold atom sample is initially produced in the higher energy F
= 2 hyperfine level. Direct absorption imaging measurements of the
peak optical depth on the $F=2\to F'=3$ transition yielded, for
this sample, $b_{o}$ $\sim$ 10 in a Gaussian radius of $r_{o}$ =
0.45 mm \cite{MOTFormulas}. This corresponds to a total number of
atoms $\sim 3.2$ $\cdot$ $10^{7}$ and a peak number density $\sim
2.2$ $\cdot$ $10^{10} atoms/cm^{3}$. However, the main goal is to
transfer the trapped atoms to a carbon-dioxide ($CO_{2}$)
laser-based optical dipole trap. The 100 W $CO_{2}$ laser operates
at 10.6 $\mu m$. The laser is focussed to a radial spot size of
$\sim$ 55 $\mu m$, and a corresponding Rayleigh range of $z_{R}$
$\sim$ 750 $\mu m$. The $CO_{2}$ laser focal zone is overlapped
with the MOT trapping region, while application of the $CO_{2}$
beam itself is controlled by a 40 MHz AOM. After the atom sample
is formed in the MOT, the $CO_{2}$ laser is applied and the sample
is compressed and loaded into the quasi static dipole trap
(QUEST).  This is done by detuning the MOT master laser 60 MHz to
the low frequency side of the trapping transition, while
simultaneously lowering the repumper intensity by an order of
magnitude.  The resulting temporal dark spot MOT loads the atoms
into the lower energy F = 1 hyperfine component, with about 15
$\%$ of the MOT atoms transferred to the QUEST. It is important to
note that this transfer efficiency is measured after a QUEST
holding period of about 1 second, during which the atomic sample
collisionally evolves towards thermal equilibrium. Measurements of
the QUEST characteristics, after the hold period, by absorption
imaging, parametric resonance, and the measured number of atoms
transferred show a sample with peak density about 6 $\cdot$
10$^{13}$ $atoms/cm^{3}$ and a temperature of 65 $\mu K$. The 1/e
lifetime of the confined atoms is greater than 5 s, limited by
background gas collisions. The residual magnetic field in the
sample area, when the MOT quadrupole field is switched off, is
estimated to be a few mG.

Here we are concerned with parametric excitation of atoms confined
to the QUEST and the sample characteristics. The sample is excited
by modulation of the $CO_{2}$ power which, for a Gaussian focused
beam, determines both axial and radial harmonic trap frequencies.
Here these are about 1.25 kHz for radial excitation and 50 Hz for
axial excitation; imposition of acoustic modulation on the
$CO_{2}$ laser AOM in the range 0 - 10 kHz is then sufficient to
drive the fundamental parametric resonances.  The characteristics
of the modulation are the modulation frequency f, the modulation
time T, and the modulation depth h. Detection of the result of
excitation is made in two ways, each based on measurements on an
image of the QUEST following a 3 ms period of free expansion. The
expansion period allows the density of the sample to be reduced
sufficiently that the QUEST becomes optically thin, so that
measurement of light scattered from the sample is proportional to
the number of atoms in the sample. In the first more traditional
method, loss of a certain number of atoms is made through
measurement of the total intensity of light scattered from the
sample.  This trap loss method measures the survival probability
of atoms in the trap.  In the second method, and the one we focus
on here, the average peak intensity in the central region of the
image is measured; this is proportional to the survival
probability of atoms more localized spatially in the harmonic
region of the trap, even after a time T of parametric resonance.
The main measured quantities are fluorescence images of the
expanded atomic cloud, these being recorded for different h, T,
and f. Characteristic results are shown in Fig. 2, where the
images show a clear increase in the axial and radial Gaussian
radii and a loss in peak intensity as T is increased at fixed h
and f = 2.5 KHz. We also draw attention to the change in shape of
the cloud with increasing T.  This is due to heating of the atom
sample while it is confined, and consequent expansion in the
weakly confining axial direction.

\begin{figure}[htpb]
\includegraphics{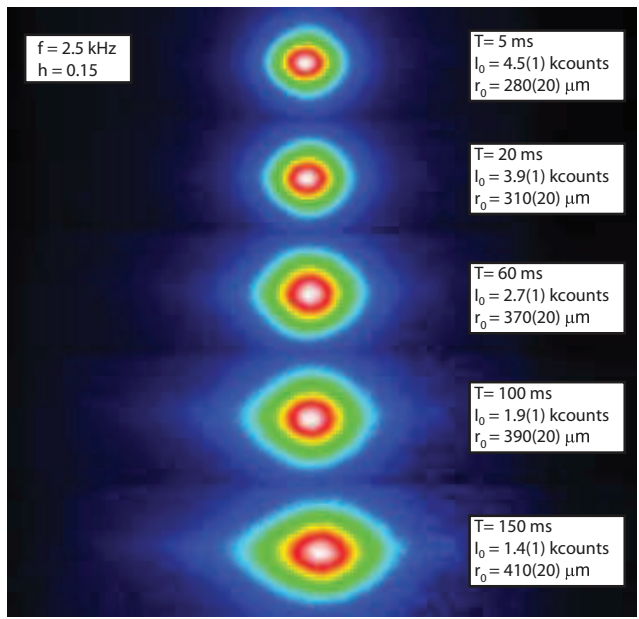}%
\caption{Typical images of the expanded cloud of ultracold
$^{87}Rb$ atoms, after a period of parametric excitation.  In all
cases the cloud is permitted to expand for 3 ms prior to imaging.
The quantity T is the modulation time, while $I_{o}$ is the peak
intensity in the central region of the image. $r_{o}$ is the
Gaussian radius of the image in the radial (vertical) direction,
which provides a spatial scale for the images.}
\label{fig2}%
\end{figure}

\begin{figure}[htpb]
\includegraphics{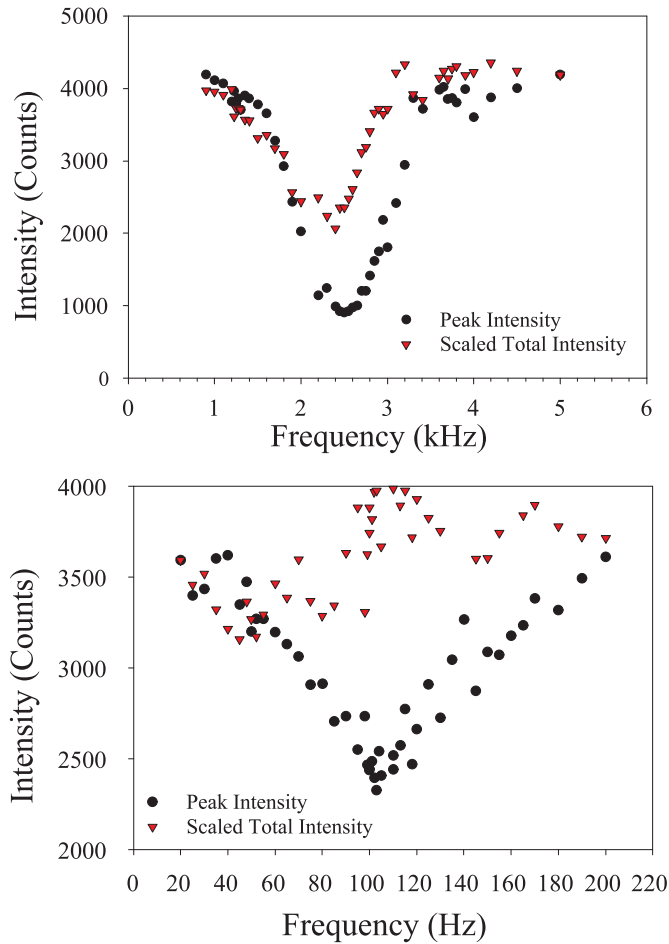}%
\caption{(a) Fundamental radial parametric resonances for h = 0.15
and T = 200 ms. (b) Fundamental axial parametric resonances for h
= 0.20 and T = 200 ms.}
\label{fig3}%
\end{figure}

We begin our analysis by showing in Fig. 3(a) the excitation
spectrum in the spectral vicinity of the fundamental radial
resonance. Two sets of data are displayed.  In the first, we show
the total number of trap atoms surviving the parametric
excitation, while in the second we show the depletion of signal
from the central spatial region of the atom cloud image.  The
differences between the two are striking. First, even for the
quite large modulation depth of 15 $\%$ the total trap loss, as a
fraction of the entire signal level, is clearly weaker than the
depletion signal. The second main feature is that the trap loss
signal is shifted by -0.3(1) KHz to lower frequencies. Both
aspects of the total trap loss signals represent significant
experimental disadvantages if one wants to characterize the lower
energy portion of the optical dipole trap potential.  We attribute
the frequency shift to the anharmonic nature of the trapping
potential; as the average amount of excitation energy of atoms in
the trap increases, the resonance harmonic frequency for those
atoms decreases, shifting the resonance to lower values. Because
the resonances can be intrinsically quite broad, this shifts the
overall response to lower frequencies. Roughly speaking, one can
envision the process as parametric evolution of an atom
distribution localized deep in the trap to a warmer one which is
more broadly distributed over the attractive part of the optical
dipole potential.  This leads to a 'freezing out' of the
parametric resonance condition and to a depletion of the number of
atoms in the deepest part of the atom trap.  We emphasize that
similar results are obtained for a range of h and T, but that the
shift of the resonance to lower frequencies becomes more
pronounced as h is increased. Within the range of our data, the
resonance positions are insensitive to T. We illustrate the point
further in Fig. 3(b), which shows the main axial resonance for the
same two measurement approaches used in Fig. 3(a), but for a
larger modulation depth of 20 $\%$. In this data, there is no
clear resonance for the trap loss signal, but a distinct one
appearing at about 105 Hz for the depletion signal. The spectral
location of this resonance is, within the experimental
uncertainty, where it is expected based on the spectral location
of the fundamental radial mode at 2.5 kHz, and assuming a trap
formed at the focus of a Gaussian trapping beam.

\begin{figure}[htpb]
\includegraphics{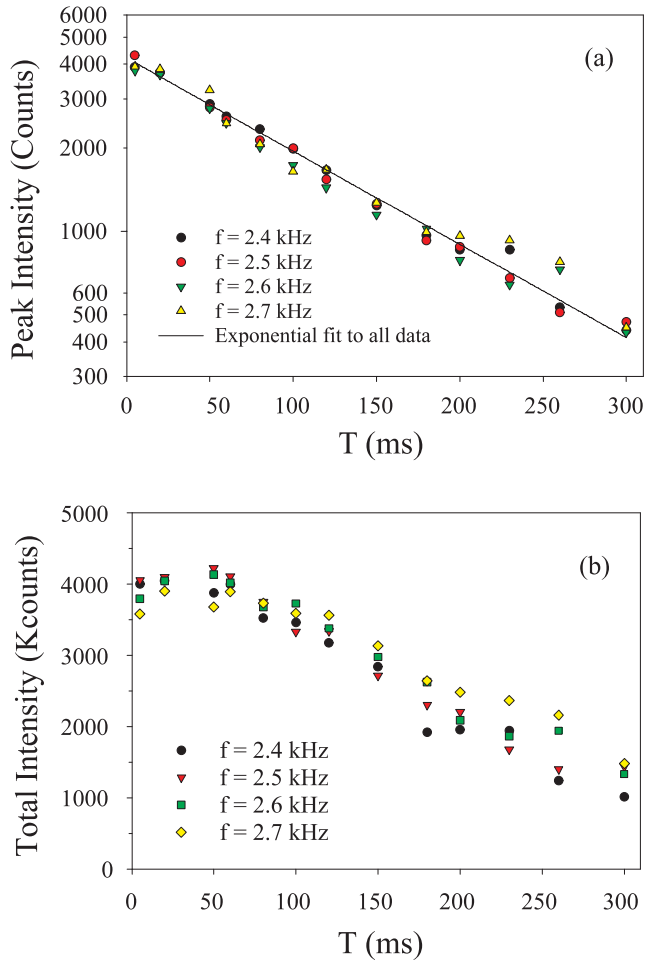}%
\caption{(a) Peak survival probability as a function of T for
several different parametric resonance excitation frequencies.
Note the exponential decay with time for longer modulation times.
(b) Relative survival probability as a function of T for several
different parametric resonance excitation frequencies.  Note the
linear decay with time for larger T. h = 0.15.}
\label{fig4}%
\end{figure}

We now turn our attention to the role of the modulation time T,
which represents the amount of time that the trap depth is
modulated before the atoms are released. Physically, T represents
the amount of time that energy may be delivered to the ultracold
atom sample.  We point out that this time scale is short compared
to the 5 s hold time of the atom sample. The variation with T of
the intensity at the peak of the image, for several modulation
frequencies, is shown in Fig. 4(a). There we see that the peak
image intensity appears to decrease exponentially for a range of
f. This decrease is evident from the smallest T = 5 ms, and
continues for a factor of 60 in T. In Fig. 4(b) the variation of
the total trap loss with T for the same modulation frequencies is
shown.  In contrast to the peak intensity, the total intensity
decreases approximately linearly with T, and further shows a
threshold below which minimal variation is measured. We
qualitatively interpret these results as follows: the atoms in the
trap are initially in thermal equilibrium at a temperature of
approximately $65 \mu K$. While the modulation is applied the
average energy of the atoms is expected to increase. Further, the
$^{87}Rb$ elastic collision rate is quite large at the
experimental conditions, so the atomic cloud is expected to stay
in approximate thermal equilibrium at some temperature.  This
temperature should increase as T increases. Even if no atoms were
removed from the trap by this process, the peak intensity would
still decrease, for both the spatial volume occupied by the atom
cloud, and the rate of free expansion are increased as temperature
is increased. However, as the trap depth is finite, some of the
atoms have sufficient energy to leave the trap.  This leads to a
decrease in the integrated image intensity (due to trap loss), and
a corresponding additional decrease in the peak image intensity.
These arguments also explain why there is a threshold in the
appearance of trap loss; the average amount of energy deposited
per atom in the system by parametric excitation must be
sufficiently large that a significant number of them have a total
energy in excess of the trap depth before trap loss is observed.
Additional measurements of the Gaussian width of the images (as in
Fig. 2) with increasing T provide a direct measure of about 800
$\mu$K/s for the heating rate, confirming this interpretation. We
also point out that there is an important relationship between the
results of Fig. 4(a) and 4(b): when the decrease in the total
number of atoms in the trap is accounted for by the data in Fig.
4(b), the remaining decrease seen in Fig. 4(a) scales
quantitatively as the mean cloud temperature $T_{c}^{3/2}$, as is
expected for the temperature dependent decrease in the peak
intensity of a Gaussian atom cloud.

Finally we point out that increasing T does not indefinitely
increase the mean temperature of the atom cloud. As predicted in
\cite{Savard,Gehm}, above a certain critical amount of energy
deposition in the system the temperature of the atom cloud remains
constant, the additional energy going into removing additional
atoms from the trap.  As we show elsewhere \cite{Balik}, this
interpretation is supported by measurements on the variation of
the Gaussian radius of the atomic cloud with T after ballistic
expansion.  The radius is seen to initially increase with T, but
then to reach a steady value corresponding to a cloud temperature
of about $35 \%$ of the trap depth, consistent with the results of
Gehm, \emph{et al.} \cite{Gehm}.

We acknowledge the financial support of the National Science
Foundation (Grant No. NSF-PHY-0654226).


\begin{thebibliography}{99}
\bibitem{MOTFormulas}For a Gaussian atom distribution of size $r_{o}$ and peak density $n_{o}$
$n(r) = n_{o}e^{-r^{2}/2r_{o}^{2}}$. The total number of atoms is
$N = (2\pi )^{3/2}n_{o}r_{o}^{3}$ and the peak optical depth is
$b_0=\sqrt{2\pi}n_0\sigma_0r_0.$ $\sigma_0$ is the weak field
resonance light scattering cross section.
\end{thebibliography}
\end{document}